\documentclass[reprint,prl,amsmath,amssymb,superscriptaddress,floatfix]{revtex4-2}
\usepackage[dvipdfmx]{graphicx}
\usepackage[dvipsnames]{xcolor}
\usepackage[colorlinks,linkcolor=Blue,citecolor=Blue,urlcolor=Blue]{hyperref}
\newcommand{\figref}[2]{\ref{#1}\hyperref[#1]{#2}}

\begin{document}
\nocite{*}

\title{Kerr-Enhanced Optical Spring}

\author{Sotatsu~Otabe}
\email[]{otabe@gw.phys.titech.ac.jp}
\affiliation{Department of Physics, Tokyo Institute of Technology, Meguro, Tokyo 152-8550, Japan}
\affiliation{Institute of Innovative Research, Tokyo Institute of Technology, Yokohama, Kanagawa 226-8503, Japan}

\author{Wataru~Usukura}
\affiliation{Department of Physics, Tokyo Institute of Technology, Meguro, Tokyo 152-8550, Japan}

\author{Kaido~Suzuki}
\affiliation{Department of Physics, Tokyo Institute of Technology, Meguro, Tokyo 152-8550, Japan}

\author{Kentaro~Komori}
\affiliation{Research Center for the Early Universe (RESCEU), Graduate School of Science, University of Tokyo, Bunkyo, Tokyo 113-0033, Japan}
\affiliation{Department of Physics, University of Tokyo, Bunkyo, Tokyo 113-0033, Japan}

\author{Yuta~Michimura}
\affiliation{LIGO Laboratory, California Institute of Technology, Pasadena, California 91125, USA}
\affiliation{Research Center for the Early Universe (RESCEU), Graduate School of Science, University of Tokyo, Bunkyo, Tokyo 113-0033, Japan}

\author{Ken-ichi~Harada}
\affiliation{Department of Physics, Tokyo Institute of Technology, Meguro, Tokyo 152-8550, Japan}

\author{Kentaro~Somiya}
\affiliation{Department of Physics, Tokyo Institute of Technology, Meguro, Tokyo 152-8550, Japan}

\date{\today}

\begin{abstract}
    We propose and experimentally demonstrate the generation of enhanced optical springs using the optical Kerr effect. A nonlinear optical crystal is inserted into a Fabry-Perot cavity with a movable mirror, and a chain of second-order nonlinear optical effects in the phase-mismatched condition induces the Kerr effect. The optical spring constant is enhanced by a factor of $1.6\pm0.1$ over linear theory. To our knowledge, this is the first realization of optomechanical coupling enhancement using a nonlinear optical effect, which has been theoretically investigated to overcome the performance limitations of linear optomechanical systems. The tunable nonlinearity of demonstrated system has a wide range of potential applications, from observing gravitational waves emitted by binary neutron star postmerger remnants to cooling macroscopic oscillators to their quantum ground state.
\end{abstract}

\maketitle
\textit{Introduction.}---In 2017, gravitational waves from a binary neutron star merger were observed for the first time~\cite{PhysRevLett.119.161101}, and electromagnetic telescopes identified its counterpart~\cite{Abbott_2017Multi}. The multimessenger observations thus realized have provided several remarkable astronomical insights, including the origin of short gamma-ray bursts~\cite{Abbott_2017GRB170817A}, synthesis of very heavy elements via the r process~\cite{Pian2017}, and novel independent measurements of the Hubble constant~\cite{Abbott2017Hubble}. However, the high-frequency gravitational waves in the $2$--$4$\,kHz band, which are predicted to be emitted by remnants possibly formed after a binary merger~\cite{Baiotti_2017,Sarin2021}, have not been observed because they lie outside the bandwidth of modern gravitational wave detectors (GWDs). Gravitational waves emitted from binary neutron star postmerger remnants contain critical information regarding high-density nuclear materials, which cannot be accessed via terrestrial experiments and is essential for determining the equation of state for neutron stars~\cite{RevModPhys.89.015007}. The internal structure exploration of neutron stars is a primary motivation for constructing third-generation GWDs~\cite{Punturo_2010,Abbott_2017CE} because postmerger signals are only observed once every few decades by second-generation GWDs~\cite{Harry_2010,Acernese_2014,Somiya_2012}. Interferometric geometry modification of second- or third-generation GWDs for specialization in the $2$--$4$\,kHz band has also been proposed~\cite{PhysRevD.99.102004,NEMO2020,PhysRevD.102.122003,PhysRevX.13.021019}.\par
Signal amplification using an optical spring is a promising technique for improving specific band sensitivity in GWDs~\cite{BRAGINSKY1999241,Buonanno_2001,PhysRevD.64.042006,PhysRevD.65.042001}. An optical spring can be generated by detuning the optical cavity to create a proportional relationship between the radiation pressure force and the test mass displacement. Under the constraint of a fixed detector bandwidth, the optical spring resonant frequency is determined by the intracavity light power~\cite{PhysRevA.69.051801,PhysRevA.74.021802,PhysRevLett.98.150802} and is limited to approximately $100$\,Hz for second-generation GWDs~\cite{PhysRevD.65.042001}. Generally, improving the impact of the optical spring is challenging because increasing intracavity power results in harmful phenomena, including thermal lensing~\cite{Harry_2010,Willke_2006} or parametric instability~\cite{PhysRevLett.95.033901,PhysRevLett.114.161102}. To address this problem, implementation of a technique called intracavity squeezing, which can expand the bandwidth of the detector~\cite{PhysRevLett.95.193001,PhysRevLett.118.143601,PhysRevX.9.011053,Korobko2019,Adya_2020}, has been investigated to increase the optical spring resonant frequency to several kHz~\cite{SOMIYA2016521,KOROBKO20182238,PhysRevD.107.122005}. This technique actively increases the signal amplification ratio of the cavity using intracavity nonlinear optical effects (NOEs) and enhances the optical spring constant without changing the intracavity power. The intracavity signal amplification method, which directly enhances optomechanical coupling, fundamentally differs from the input squeezing technique~\cite{PhysRevD.23.1693,PhysRevD.65.022002,PhysRevLett.124.171101,PhysRevLett.124.171102,PhysRevLett.131.041403} currently used for practical GWDs~\cite{PhysRevLett.123.231107,PhysRevLett.123.231108}, and these techniques can be used in combination~\cite{PhysRevLett.131.143603}.\par
It is worth noting that enhanced optomechanical coupling has been extensively researched beyond the context of GWDs. Intracavity squeezing has the potential to cool an optomechanical oscillator to the ground state in the unresolved sideband regime~\cite{PhysRevA.79.013821,Asjad:19,https://doi.org/10.1002/lpor.201900120}, which closely relates to enhancing optical damping (i.e., the imaginary component of the complex optical spring constant) using NOEs. Moreover, the application of enhanced optomechanical coupling via intracavity NOEs to generate strong mechanical squeezing~\cite{PhysRevA.93.043844,https://doi.org/10.1002/andp.201900596,Zhang:20}, manipulate normal-mode splitting~\cite{PhysRevA.80.033807,PhysRevA.81.013835}, and realize single-photon quantum processes~\cite{PhysRevLett.114.093602,Wang:20} has been theoretically investigated. Direct observation of enhanced optomechanical coupling represents a significant step toward achieving the full potential of future optomechanical systems.\par
This Letter reports on observations from an experimental demonstration focused on optical spring enhancement via NOEs, which is essential for intracavity signal amplification systems. We induced the NOE in a Fabry-Perot type optomechanical cavity. Among the various candidates for NOEs capable of generating squeezed states~\cite{SCHNABEL20171}, our research revealed that the optical Kerr effect is a promising approach. This effect provides significant signal amplification effects while maintaining sufficient intracavity power, and thus, strong optomechanical coupling can be generated.\par
\begin{figure}
	\centering
	\includegraphics[width=\hsize]{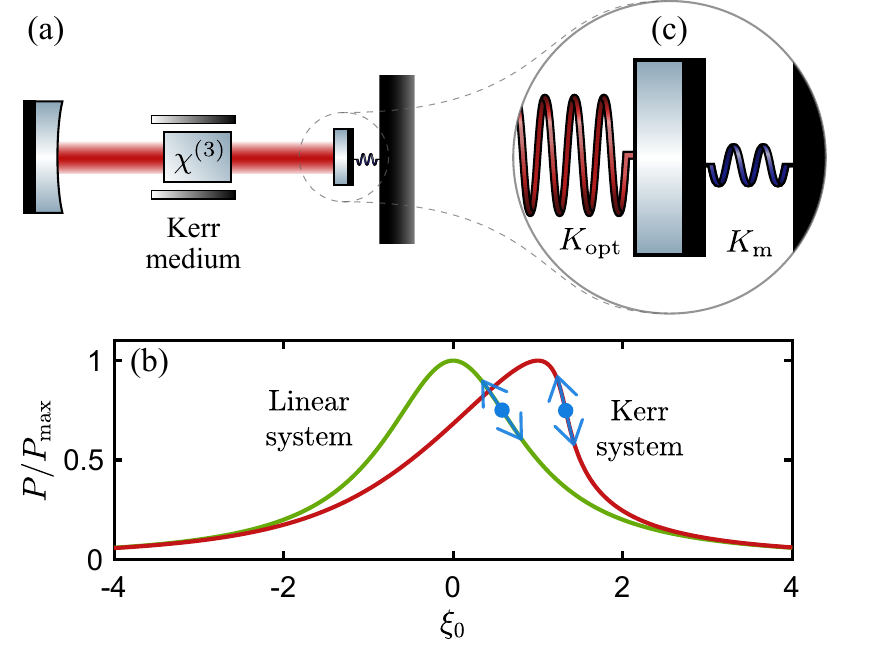}
	\caption{(a) Schematic of an optomechanical system whose coupling is enhanced by the optical Kerr effect. An optical crystal with third-order nonlinearity $\chi^{(3)}$ (Kerr medium) is inserted into a laser-driven optical cavity with a movable-end mirror. (b) Theoretical intracavity power curve. Linear and Kerr systems correspond to the cases of Kerr gains at $\zeta=0$ and $\zeta=-1$, respectively. The vertical axis is normalized by the intracavity power at resonance $P_\mathrm{max}=2\mathcal{F}P_0/\pi$. (c) Schematic of the mechanical and enhanced optical springs, whose complex spring constants are $K_\mathrm{m}$ and $K_\mathrm{opt}$, respectively. The positive optical spring resulting from the radiation pressure restoring force is enhanced by signal amplification due to the Kerr effect with $\zeta<0$.}
	\label{fig:Kerr-outline}
\end{figure}
\textit{Kerr-enhanced optomechanical system.}---We investigate a system wherein the Kerr medium is inserted into the optomechanical cavity [Fig.\,\figref{fig:Kerr-outline}{(a)}]. The dynamics of this system can be described as the sum of the optomechanical Hamiltonian~\cite{PhysRevA.51.2537,RevModPhys.86.1391} and Kerr Hamiltonian $\hat{H}_\mathrm{Kerr}=(\hbar/2)\chi(\hat{a}^\dagger)^2\hat{a}^2$~\cite{PDDrummond_1980}. Here, $\hbar$ is the reduced Planck constant, $\chi$ is the quantity proportional to the nonlinear susceptibility of the Kerr medium, and $\hat{a}^\dagger$ and $\hat{a}$ are the creation and annihilation operators of the optical mode, respectively. The differential equation for the complex light-field amplitude $a(t)$ in the cavity without intracavity losses can be written as
\begin{equation}
	\dot{a}=\left[i\Delta'+iG x-i\chi n-\gamma\right]a+\sqrt{2\gamma} a_\mathrm{in},
\end{equation}
where $x$ is the sum of the mechanical and photothermal displacements~\cite{Ma:21,Otabe:22}, $G$ is the optomechanical frequency shift per displacement, and $a_\mathrm{in}$ is the amplitude of the drive laser. The Kerr effect practically changes the angular frequency of cavity detuning $\Delta'$ proportionally to the intracavity photon number $n(t)=|a(t)|^2$, but it does not affect the cavity decay rate $\gamma$.\par
Let us make a linear approximation as $a(t)=\bar{a}+\delta a(t)$ and $x(t)=\bar{x}+\delta x(t)$ to examine the behavior of the light-field amplitude around a stable point. The intracavity power $P$ can be defined by the average number of photons inside the cavity $\bar{n}=|\bar{a}|^2$, calculated from the zero-order terms as
\begin{equation}\label{eq:Kerr-P}
	P=\frac{\hbar\omega_0c}{2L}\bar{n}=\frac{2\mathcal{F}}{\pi}\frac{1}{1+\xi^2}P_0,
\end{equation}
where $\omega_0$ is the angular frequency of the carrier light, $c$ is the speed of light, $L$ is the half cycle length of the cavity, $\mathcal{F}$ is the cavity finesse, and $P_0$ is the incident power of the carrier light on the cavity. The normalized cavity detuning of the Kerr system $\xi=\xi_0+\xi_\mathrm{K}$ consists of two components, where $\gamma\xi_0=\Delta'+G \bar{x}$ is the shifted cavity detuning of the linear optomechanical system, and $\gamma\xi_\mathrm{K}=-\chi\bar{n}$ is the cavity detuning derived from the refractive index change due to the Kerr effect. The asymmetry of the cavity spectrum of the Kerr system [Fig.\,\figref{fig:Kerr-outline}{(b)}] is a consequence of the nonlinear phase shift, which is proportional to the intracavity power.\par
The motion of a movable mirror in the optomechanical cavity is dominated by the radiation pressure force modified by the Kerr effect. We focus on the optical spring, a phenomenon resulting from the proportionality of the radiation pressure force to the displacement of the test mass. The optical spring constant in a linear system is proportional to the product of the intracavity power and cavity finesse, which generally has technical limits. In contrast, in the Kerr system, a drastic gradient of the radiation pressure force exists compared to a linear system. Therefore, we can generate an optical spring enhanced by the Kerr effect without changing the intracavity power or the cavity linewidth [Fig.\,\figref{fig:Kerr-outline}{(c)}]. The complex optical spring constant for the angular frequency $\Omega$ is obtained from simultaneous equations for the first-order micro terms of the light-field amplitude and cavity length, which can be written as~\cite{Kerrsupp}
\begin{equation}\label{eq:Kerr-Kopt}
	K_\mathrm{opt}(\Omega)=\frac{4\omega_0 P}{Lc\gamma}\dfrac{\xi}{(1+i\Omega/\gamma)^2+\xi^2+2\xi\xi_\mathrm{K}}.
\end{equation}
The real and imaginary components of Eq.\,\eqref{eq:Kerr-Kopt} correspond to the optical spring and damping constant, respectively. The dimensionless Kerr gain $\zeta$, which does not depend on cavity detuning, follows from Eq.\,\eqref{eq:Kerr-P} as
\begin{equation}
	\zeta=(1+\xi^2)\xi_K=-\dfrac{2\chi P_0}{\gamma^2\hbar\omega_0}.
\end{equation}
When $\zeta$ exceeds a threshold value of $\zeta_0=-8/(3\sqrt{3})\sim -1.54$, multiple intracavity powers correspond to a single cavity detuning, indicating that the cavity response enters a multistable state~\cite{AGWhite_1996,PhysRevLett.95.193001,PhysRevA.80.053801,PhysRevA.84.033839}. The signal amplification ratio due to the Kerr effect increases by approaching a multistable regime.\par
\begin{figure}
	\centering
	\includegraphics[width=\hsize]{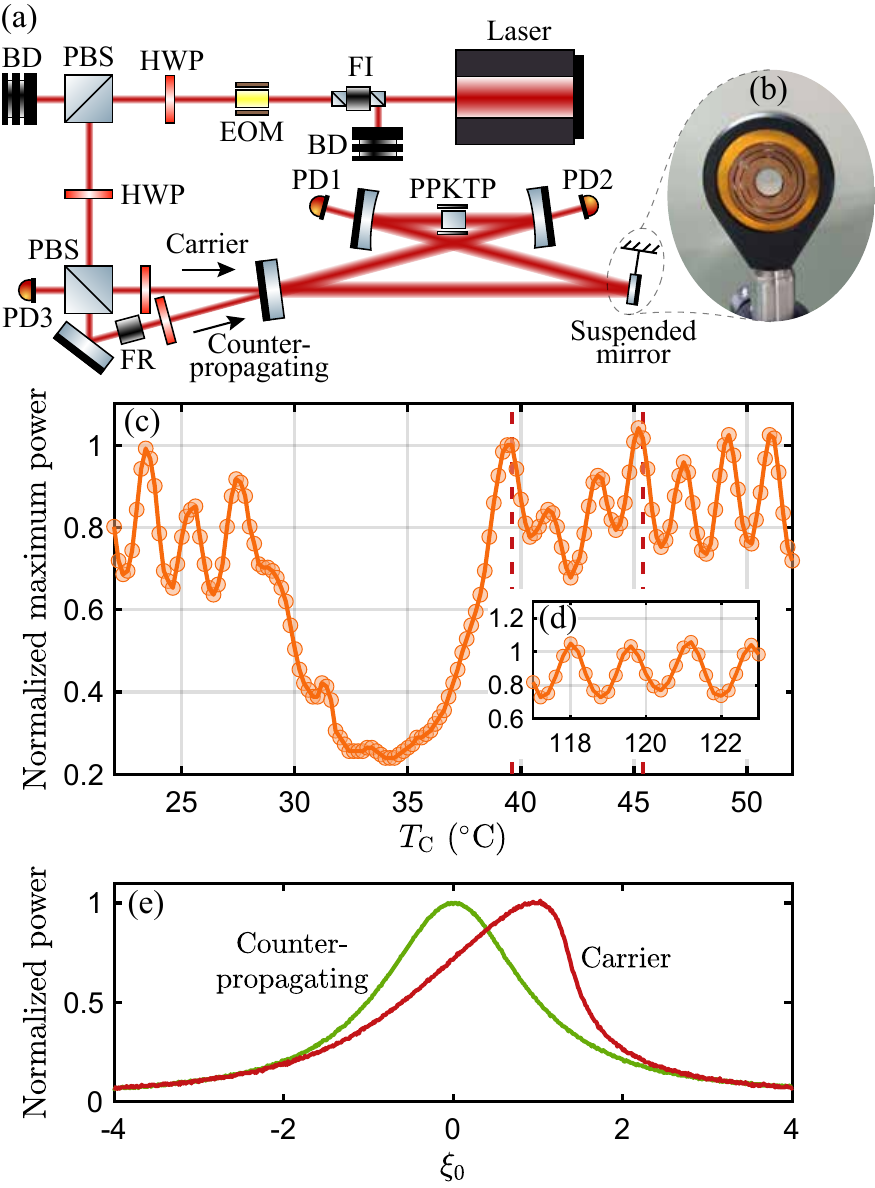}
	\caption{(a) Experimental setup. Light emitted from a $1064$\,nm Nd:YAG laser passed through a Faraday isolator (FI) and an electro-optic modulator (EOM) and was attenuated by a half-wave plate (HWP), polarizing beam splitter (PBS), and beam damper (BD). A periodically poled KTiOPO$_4$ (PPKTP) crystal was inserted into the cavity to induce NOEs. (b) Image of a small mirror suspended by a double spiral spring. A magnet was attached to the back of the mirror, and the cavity length was varied through a coil magnet actuator. (c),(d) Transmitted light power at resonance versus crystal temperature $T_\mathrm{C}$. The vertical axes are normalized by the transmitted light power at resonance in the first phase-mismatched condition ($39.6$\,$^\circ$C). The red dotted lines indicate the setting temperature for the optical spring constant measurement. The temperatures producing the phase-mismatched conditions remain periodically present in the high-temperature region, as shown in (d). (e) Measured cavity spectra. The horizontal axis was normalized by the half width at half maximum of the counterpropagating light power.}
	\label{fig:Kerr-setup}
\end{figure}
\begin{figure*}
	\centering
	\includegraphics[width=\hsize]{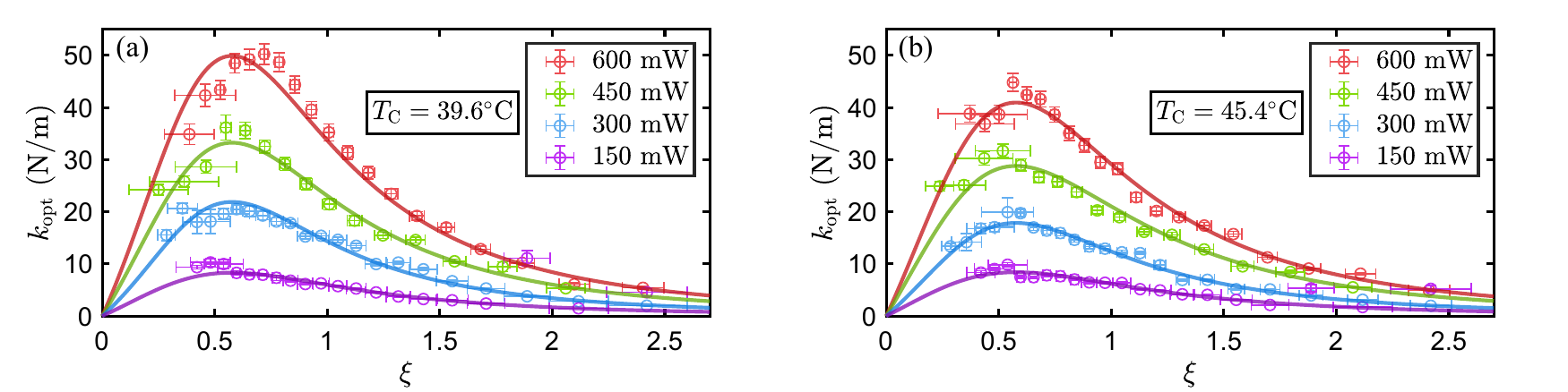}
	\caption{(a),(b) Estimation of the optical spring constant $k_\mathrm{opt}=K_\mathrm{opt}(0)$. The input power $P_0$ was set to $600$, $450$, $300$, and $150$\,mW, and the cavity detuning $\xi$ was finely varied. The crystal temperature $T_\mathrm{C}$ was $39.6$\,$^\circ$C for (a) and $45.4$\,$^\circ$C for (b). The circles with error bars represent the estimated optical spring constants obtained from the transfer function measurements, and the solid lines in each color show the fitting results with the Kerr gain $\zeta$ and the maximum value of the optical spring constant for the linear theory $k_\mathrm{opt-0}=K_\mathrm{opt}(0)|_{\zeta=0,\xi=1/\sqrt{3}}$ as parameters.}
	\label{fig:Kerr-kopt}
\end{figure*}
\textit{Experimental setup.}---We used a bow-tie cavity containing a $10$-mm-long nonlinear optical crystal at a beam waist with a radius of $40$\,\textmu m [Fig.\,\figref{fig:Kerr-setup}{(a)}]. The measured finesse of this cavity is $\mathcal F\simeq 100\pm10$, which agrees with the design value. One of the mirrors constituting the cavity was small with a weight of $280$\,mg and diameter of $6.35$\,mm, suspended by a double spiral spring [Fig.\,\figref{fig:Kerr-setup}{(b)}] with a resonant frequency of $14.0\pm0.1$\,Hz and mechanical Q factor of $193\pm3$. This mirror was also coupled to an optical spring, allowing the precise measurement of the resonant frequencies of the composite spring.\par
NOEs were induced in phase-mismatched conditions. Figures\,\figref{fig:Kerr-setup}{(c)} and \figref{fig:Kerr-setup}{(d)} shows the dependence of the maximum transmitted power of the carrier light on the crystal temperature. At the phase-matching temperature ($34.2$\,$^\circ$C), the majority of the carrier light was converted into the second harmonic wave, decreasing the transmitted power. When the crystal temperature exceeded this temperature, almost no second harmonic was generated at some temperatures (e.g., $39.6$\,$^\circ$C), and this condition was denoted as a phase mismatch. The second harmonic generation (SHG) processes are induced even in the phase-mismatched condition. However, the phases of the second harmonics were misaligned at each location in the crystal because the second harmonics propagate at a different speed than the fundamental wave. The generated second harmonics were canceled and reconverted into the fundamental wave with a different phase from the original. A chain of second-order NOEs induces a phase shift of the fundamental wave proportional to incident intensity, which is equivalent to the optical Kerr effect~\cite{PhysRevLett.74.2816,GeorgeIStegeman_1997}. The effective velocity difference between the fundamental and second harmonics, and thus the effective nonlinear susceptibility $\chi$, depends on the crystal temperature. At temperatures slightly different from other phase-mismatched conditions (e.g., $45.4$\,$^\circ$C), the intracavity loss due to the SHG effect was equivalent to that of the first phase-mismatched condition. We compared the difference in optical spring constants for different Kerr gains by measuring them at temperatures with similar intracavity losses.\par
We injected light bidirectionally into the cavity. The transmitted carrier and counterpropagating light were measured using photodetectors (PDs) 1 and 2, and a Faraday rotator (FR) was used to measure the reflected carrier light with PD3. Figure\,\figref{fig:Kerr-setup}{(e)} shows the transmitted light power of the carrier and the counterpropagating light under the first phase-mismatched condition, with input powers of $530$ and $50$\,mW, respectively. Although no essential difference exists between these two paths, we can examine the impact of the Kerr effect by injecting light of approximately $1$ order of magnitude different powers into each path. The counterpropagating light is hardly affected by the Kerr effect and shows a Lorentzian curve, while the carrier light spectrum indicates that a nonlinear refractive index has been induced. This measurement was performed by inserting the crystal at an inclination relative to the cavity axis to avoid coupling of the carrier and counterpropagating light due to stray light on the antireflection coating. In this case, the first phase-mismatch temperature and Kerr gain changed. The crystal was reinstalled parallel to the cavity axis in subsequent measurements to obtain the highest possible Kerr gain.\par
\textit{Results.}---We estimated the optical spring constant by applying a force to the test mass and then measuring the change in the cavity length from the reflected and transmitted light of the carrier path. The cavity length does not directly correspond to the displacement of the test mass owing to the photothermal effect in the nonlinear optical crystal. In our experimental system, it is critical to accurately measure the photothermal displacement, which rotates the quadrature of the optical spring~\cite{PhysRevD.92.062003,Altin2017,Otabe:22}. Photothermal parameters can be estimated independently and highly accurately in an experimental setup not affected by an optical spring~\cite{Ma:21}. To compensate for the photothermal effect, we measured the photothermal displacement using a setup where a mirror with a piezoelectric element replaced the suspended mirror for each parameter used in the optical spring estimation process~\cite{Kerrsupp}.\par
Figures\,\figref{fig:Kerr-kopt}{(a)} and \figref{fig:Kerr-kopt}{(b)} show the estimated results of the optical spring constant. The measured optical spring constants, for the same cavity detuning and input power, were larger when $T_\mathrm{C}=39.6$\,$^\circ$C, and a difference was observed at a higher input light power, reflecting that the Kerr gain was larger at $39.6$\,$^\circ$C and proportional to the input light power. Slight systematic errors were expected due to changes in phase mismatching caused by thermal absorption in the nonlinear optical crystal~\cite{Kerrsupp}. Although the additional SHG loss was larger at $39.6$\,$^\circ$C, the measured optical spring constant was also larger. Conclusively, the increase in optical spring constant owing to the Kerr effect was more significant than the decrease due to the SHG loss.\par
\begin{figure}
	\centering
	\includegraphics[width=\hsize]{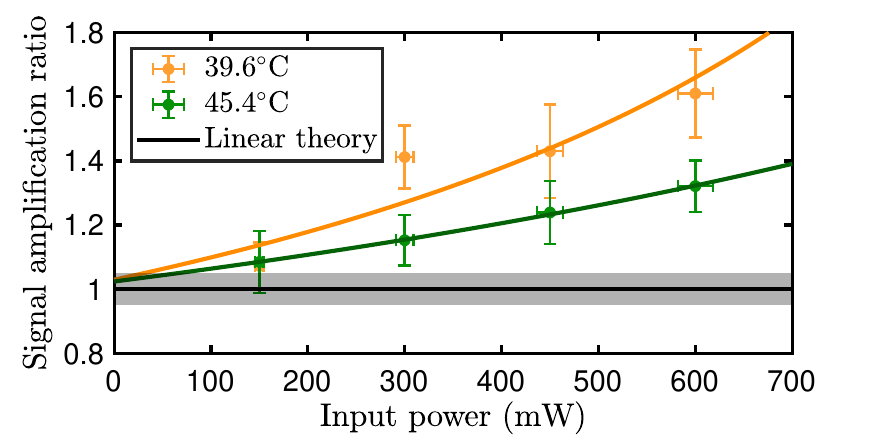}
	\caption{Net signal amplification ratio obtained by normalizing the estimated maximum value of the optical spring constant using the relative input light power. The filled circles with error bars represent the estimated signal amplification ratio, and the solid lines in each color show the fitting with the Kerr theory. The vertical axis is normalized by $k_\mathrm{opt-0}$ estimated for $T_\mathrm{C}=39.6$\,$^\circ$C and $P_0=600$\,mW, and the gray-filled area corresponds to its estimated error.}
	\label{fig:Kerr-ampratio}
\end{figure}
The signal amplification induced by the Kerr effect depends on the cavity detuning, and the difference from the linear system is maximized for $\xi=1/\sqrt{3}$. This condition also provides the maximum value of the optical spring constant for both the Kerr and linear systems. Figure\,\ref{fig:Kerr-ampratio} shows the signal amplification ratio, i.e., the enhancement factor of the maximum optical spring constant compared to the linear theory. The estimated values for each input light power show clear differences between the two temperatures, indicating that the effective nonlinear susceptibility $\chi$ is adjustable. Conclusively, the optical spring is enhanced by the optical Kerr effect with a signal amplification ratio of up to $1.6\pm0.1$.\par
\textit{Discussion.}---The experiment was based on the intracavity signal amplification method via an optical Kerr effect, which produced nontrivial amplification effects that could not be obtained by simply increasing the intracavity power or cavity finesse. The Kerr effect, estimated from optical spring measurements, was up to $-1.9\pm0.2\times10^{-17}$\,$\mathrm{m}^2/\mathrm{W}$ in terms of the nonlinear refractive index. Our result was larger than the value of $-2\times10^{-18}$\,$\mathrm{m}^2/\mathrm{W}$ obtained in the previous study~\cite{PhysRevA.80.053801} because the periodic poling allowed the use of a larger nonlinear susceptibility~\cite{Shoji2002}. In contrast, the nonlinear refractive index derived from the third-order nonlinearity of KTP was approximately $3\times10^{-19}$\,$\mathrm{m}^2/\mathrm{W}$~\cite{LI199775}, sufficiently smaller than the estimation error. The third-order nonlinearity could be observed with more precise measurements as a nonlinear refractive index independent of phase mismatching.\par
Further increase in the input light power could lead to multistability and considerably larger optical spring constants. From the fitting, the critical incident powers at which the optical spring constant diverges were predicted as $1.56\pm0.37$\,W for $T_\mathrm{C}=39.6$\,$^\circ$C and $2.65\pm0.05$\,W for $T_\mathrm{C}=45.4$\,$^\circ$C. These conditions were sufficiently feasible, and we observed hysteresis in the spectrum, suggesting multistability, by replacing the input coupler and increasing the finesse to approximately $300$~\cite{Kerrsupp}. A higher signal amplification ratio would be observed by resolving the control instability.\par
In third-order NOEs, the Kerr gain is determined by the intracavity intensity and nonlinear susceptibility of the crystal. By inducing a chain of second-order NOEs, as in our experimental system, a variable Kerr gain can be achieved by varying the phase mismatch. Furthermore, the same scheme can be used to enhance backaction cooling in regions of $\xi<0$ because the sign of the Kerr effect can be switched by decreasing the crystal temperature below the phase-matching condition~\cite{PhysRevLett.74.2816,GeorgeIStegeman_1997}.\par
Although experimental attempts have previously been made to enhance optomechanical coupling with optical parametric amplification~\cite{otabethesis,10.1063/5.0137001}, significant differences with the linear system were not observed because the SHG becomes dominant owing to the high intracavity power required to generate the optical spring. This property contrasts the Kerr scheme, wherein the effective signal amplification ratio can be increased with a higher carrier intensity. Moreover, the Kerr scheme is relatively easy to implement because this scheme does not add degrees of freedom that must be controlled. For GWDs, Kerr media shall be inserted into arm cavities with large beam size and high intracavity power~\cite{PhysRevLett.95.193001}. The correspondence between the optical parametric amplification and Kerr schemes is detailed in the Supplemental Material~\cite{Kerrsupp} (see also references~\cite{Drever1983,PhysRevA.31.3068} therein).\par
\textit{Conclusion.}---We constructed an intracavity signal amplification system based on the optical Kerr effect and succeeded in enhancing the optical spring constant by a factor of $1.6\pm0.1$. Our results correspond to an increase in the resonant frequency of the optical spring from $53\pm 1$\,Hz to $67\pm 3$\,Hz. A more significant signal amplification ratio could be achieved by approaching the multistable state and solving technical problems. The proposed scheme is easy to implement and provides a novel tunable parameter for optomechanical systems. We anticipate it will play a key role in the quantum manipulation of macroscopic optomechanical systems such as GWDs.\par
This work was supported by JST CREST (JPMJCR1873), Grant-in-Aid for JSPS Fellows (20J22778), and the Precise Measurement Technology Promotion Foundation. We would like to thank Jerome Degallaix and colleagues at LMA for providing us with a specially coated mirror and John Winterflood at UWA for designing the double spiral spring.

\bibliography{Kerrbib}

\onecolumngrid
\newpage

\begin{center}
\textbf{\large Supplementary Materials for ``Kerr-Enhanced Optical Spring''}
\end{center}

\setcounter{equation}{0}
\setcounter{figure}{0}
\setcounter{table}{0}
\setcounter{page}{1}
\makeatletter
\renewcommand{\theequation}{S\arabic{equation}}
\renewcommand{\thefigure}{S\arabic{figure}}

\section{Derivation of the Kerr-enhanced optical spring constant}
This section describes the properties of an optomechanical system enhanced by the optical Kerr effect based on the Hamiltonian formulation~\cite{PhysRevA.51.2537,RevModPhys.86.1391}. To describe this system, the Kerr-Hamiltonian $\hat{H}_\mathrm{Kerr}$~\cite{PDDrummond_1980} induced by third-order nonlinear polarization must be considered:
\begin{equation}
	\hat{H}_\mathrm{Kerr}=\dfrac{\hbar}{2}\chi(\hat{a}^\dagger)^2\hat{a}^2,
\end{equation}
where $\hbar$ is the reduced Planck constant, $\chi$ is the quantity proportional to the nonlinear susceptibility of the Kerr medium, and $\hat{a}^\dagger$ and $\hat{a}$ are the creation and annihilation operators of the optical mode, respectively. The term added to the quantum Langevin equation is calculated as
\begin{equation}
	-\dfrac{i}{\hbar}\left[\hat{a},\hat{H}_\mathrm{Kerr}\right]=-i\chi\hat{a}^\dagger\hat{a}^2.
\end{equation}
Here, we assume that the photon number is sufficiently large and describe the stochastic differential equation for the complex light-field amplitude $a(t)=\left<\hat{a}(t)\right>$:
\begin{equation}
	\dot{a}=\left[i\Delta'+iGx-i\chi n-\gamma'-\beta n\right]a+\sqrt{2\gamma_\mathrm{in}} a_\mathrm{in}, 
\end{equation}
where $\Delta'$ is the angular frequency of the cavity detuning, $G$ is the optomechanical frequency shift per displacement, and $a_\mathrm{in}$ is the amplitude of the drive laser. The cavity decay rate is divided into two terms, $\gamma'=\gamma_\mathrm{in}+\gamma_\mathrm{out}$ and $\beta n$, where the latter corresponds to the low-efficiency second harmonic generation (SHG) loss resulting from the deviation from the ideal phase-mismatched condition. Here, $\gamma_\mathrm{in}$ and $\gamma_\mathrm{out}$ are the decay rates of the input coupler and intracavity losses excluding the SHG loss, respectively. The decay rate of the low-efficiency SHG process is proportional to the quantity corresponding to the number operator $n=|a|^2$ with a proportionality factor of $\beta$. The effective cavity length $x$ is equal to the sum of the displacement of the test mass $x_\mathrm{act}$ and photothermal displacement $x_\mathrm{th}$~\cite{Ma:21,Otabe:22}. The differential equation for $x_\mathrm{th}$ can be written as
\begin{equation}
	\dot{x}_\mathrm{th}=-\gamma_\mathrm{th}x_\mathrm{th}+d\hbar Gn,
\end{equation}
where $\gamma_\mathrm{th}=1/(kC)$ is the photothermal relaxation rate and $d=\alpha\alpha'L'^2c/(2C)$ is the proportionality coefficient for heat absorption. $k$ is the thermal resistance, $C$ is the heat capacity, $\alpha$ is the coefficient of linear thermal expansion, $\alpha'$ is the absorption coefficient, $L'$ is the crystal length, and $c$ is the speed of light. The equation of motion for $x_\mathrm{act}$ can be written as
\begin{equation}
	m\ddot{x}_\mathrm{act}=-m\Omega_\mathrm{m}^2x_\mathrm{act}-m\Gamma_\mathrm{m}\dot{x}_\mathrm{act}+\hbar G n+F_\mathrm{ext},
\end{equation}
where $m$ is the mirror mass, $\Omega_\mathrm{m}$ is the mechanical resonant angular frequency, $\Gamma_\mathrm{m}$ is the mechanical damping constant, and $F_\mathrm{ext}$ is the external force applied to the test mass. The third term on the right-hand side represents the radiation pressure force.\par
For a stable point where $\dot{x}_\mathrm{act}=0$, $\dot{x}_\mathrm{th}=0$, and $\dot{a}=0$, we  can make linear approximations as $x_\mathrm{act}(t)=\bar{x}_\mathrm{act}+\delta x_\mathrm{act}(t)$, $x_\mathrm{th}(t)=\bar{x}_\mathrm{th}+\delta x_\mathrm{th}(t)$, $F_\mathrm{ext}(t)=\bar{F}_\mathrm{ext}+\delta F_\mathrm{ext}(t)$, and $a(t)=\bar{a}+\delta a(t)$. Further, $\bar{x}=\bar{x}_\mathrm{act}+\bar{x}_\mathrm{th}$, $\delta x=\delta x_\mathrm{act}+\delta x_\mathrm{th}$, $\bar{n}=|\bar{a}|^2$, and $\delta n=\bar{a}^*\delta a+\bar{a}\delta a^*$. Here, we neglect the classical and quantum fluctuations in the input field: $a_\mathrm{in}=\bar{a}_\mathrm{in}$. The intracavity carrier light power $P$ can be calculated from the time average of the light-field amplitudes as
\begin{equation}
	P=\dfrac{\hbar\omega_0 c}{2L}\bar{n}=\dfrac{c}{L}\dfrac{\gamma_\mathrm{in}}{\gamma^2+\Delta^2}P_0,
\end{equation} 
where $\omega_0\simeq GL$ is the angular frequency of the carrier light, $L$ is the half-cycle length of the cavity, $\gamma=\gamma'+\beta\bar{n}$ and $\Delta=\Delta'+G\bar{x}-\chi\bar{n}$ are the effective cavity decay rate and detuning, respectively, and $P_0=\hbar\omega_0|\bar{a}_\mathrm{in}|^2$ is the incident power of the carrier light on the cavity. The simultaneous differential equations for the first-order micro-terms are written as
\begin{eqnarray}
	\delta\dot{a}&=&[i\Delta-\gamma]\delta a+(iG\delta x-i\chi\delta n-\beta\delta n)\bar{a}\label{eq:Kerrsupp-DiffEqfora},\\
	\delta\dot{x}_\mathrm{th}&=&-\gamma_\mathrm{th}\delta x_\mathrm{th}+d\hbar G\delta n\label{eq:Kerrsupp-DiffEqforxth},\\
	\delta\ddot{x}_\mathrm{act}&=&-\Omega_\mathrm{m}^2\delta x_\mathrm{act}-\Gamma_\mathrm{m}\delta \dot{x}_\mathrm{act}+\hbar G\delta n/m+\delta F_\mathrm{ext}/m.\label{eq:Kerrsupp-DiffEqforxact}
\end{eqnarray}
The Fourier transform of Eq.\,\eqref{eq:Kerrsupp-DiffEqfora} yields
\begin{equation}
	\delta a(\Omega)=\chi_\mathrm{c}(\Omega)(iG\delta x(\Omega)-i\chi\delta n(\Omega)-\beta\delta n(\Omega))\bar{a},
\end{equation}
where $\chi_\mathrm{c}=1/(i\Omega-i\Delta+\gamma)$ denotes the susceptibility of the cavity. Furthermore, because $(\delta a^*)(\Omega)=(\delta a(-\Omega))^*$, we obtain
\begin{equation}
	(\delta a^*)(\Omega)=\chi_\mathrm{c}^*(-\Omega)(-iG\delta x(\Omega)+i\chi\delta n(\Omega)-\beta\delta n(\Omega))\bar{a}^*.
\end{equation}
We use $(\delta x(-\Omega))^*=\delta x(\Omega)$ and $(\delta n(-\Omega))^*=\delta n(\Omega)$ because $\delta x(t)$ and $\delta n(t)$ are real numbers. The photon number fluctuation $\delta n(\Omega)$ is calculated as follows:
\begin{eqnarray}\label{eq:Kerrsupp-dndx}
	\delta n(\Omega)&=&i\bar{n}\left[\chi_\mathrm{c}(\Omega)-\chi_\mathrm{c}^*(-\Omega)\right](G\delta x(\Omega)-\chi \delta n(\Omega))-\beta\bar{n}\left[\chi_\mathrm{c}(\Omega)+\chi_\mathrm{c}^*(-\Omega)\right]\delta n(\Omega)\nonumber\\
	&=&-2G\bar{n}\dfrac{\Delta}{(\gamma+i\Omega)^2+\Delta^2+2[\beta(\gamma+i\Omega)-\chi\Delta]\bar{n}}\delta x(\Omega).
\end{eqnarray}
This indicates that the Kerr effect enhances the response of the photothermal backaction and the radiation pressure force for the cavity length. Substituting Eq.\,\eqref{eq:Kerrsupp-dndx} into Eq.\,\eqref{eq:Kerrsupp-DiffEqforxth}, we obtain the influence of the photothermal effect on the effective cavity length:
\begin{equation}
	(i\Omega+\gamma_\mathrm{th})\delta x_\mathrm{th}(\Omega)=-\omega_\mathrm{th}(\Omega)\delta x(\Omega),\label{eq:Kerrsupp-dxthdx}
\end{equation}
where
\begin{equation}
	\omega_\mathrm{th}(\Omega)=2d\hbar G^2\bar{n}\dfrac{\Delta}{(\gamma+i\Omega)^2+\Delta^2+2[\beta(\gamma+i\Omega)-\chi\Delta]\bar{n}}\label{eq:Kerrsupp-PTAbsConst}
\end{equation}
denotes the photothermal absorption rate. Substituting Eq.\,\eqref{eq:Kerrsupp-dndx} into Eq.\,\eqref{eq:Kerrsupp-DiffEqforxact}, the susceptibility of the optomechanical oscillator can be calculated as
\begin{equation}
	\dfrac{\delta x_\mathrm{act}(\Omega)}{\delta F_\mathrm{ext}(\Omega)}=\dfrac{1}{m(-\Omega^2+\Omega^2_\mathrm{m}+i\Omega\Gamma_\mathrm{m})+\Sigma_\mathrm{th}(\Omega)},
\end{equation}
where $\Sigma_\mathrm{th}(\Omega)$ is the optomechanical self-energy defined as
\begin{equation}
	\Sigma_\mathrm{th}(\Omega)=\dfrac{\gamma_\mathrm{th}+i\Omega}{(\omega_\mathrm{th}(\Omega)+\gamma_\mathrm{th})+i\Omega}K_\mathrm{opt}(\Omega),
\end{equation}
and the complex optical spring constant $K_\mathrm{opt}(\Omega)$ is calculated as
\begin{equation}
	K_\mathrm{opt}(\Omega)=2\hbar G^2\bar{n}\dfrac{\Delta}{(\gamma+i\Omega)^2+\Delta^2+2[\beta(\gamma+i\Omega)-\chi\Delta]\bar{n}}.\label{eq:Kerrsupp-CKopt}
\end{equation}
The real and imaginary components in Eq.\,\eqref{eq:Kerrsupp-CKopt} correspond to the optical spring and damping constants, respectively. If the measurement bandwidth is sufficiently smaller than the cavity decay rate ($\Omega\ll\gamma$), the optical spring constant $k_\mathrm{opt}$ can be written as
\begin{equation}\label{eq:Kerrsupp-kopt}
	k_\mathrm{opt}\simeq K_\mathrm{opt}(0)=\dfrac{4\omega_0 P}{Lc}\dfrac{\Delta}{\gamma^2+\Delta^2+2\gamma\gamma_\mathrm{S}+2\Delta\Delta_\mathrm{K}},
\end{equation}
where we define the cavity decay rate owing to SHG loss and cavity detuning owing to the Kerr effect as $\gamma_\mathrm{S}=\beta\bar{n}$ and $\Delta_\mathrm{K}=-\chi\bar{n}$, respectively.\par
Even in the presence of SHG loss, a significant signal amplification effect can be obtained if a cavity detuning $\Delta$ exists for which the denominator in Eq.\,\eqref{eq:Kerrsupp-kopt} equals zero. The discriminant for variable $\Delta$ is obtained as
\begin{equation}
	D(\bar{n})=(\chi^2-3\beta^2)\bar{n}^2-4(\gamma_\mathrm{in}+\gamma_\mathrm{out})\beta\bar{n}-(\gamma_\mathrm{in}+\gamma_\mathrm{out})^2.
\end{equation}
For the solution of the equation $D(\bar{n})=0$ to be positive, $D(\bar{n})$ must be a convex downward function because $D(0)<0$. Therefore, the signal amplification induced by the Kerr effect is more pronounced than the reduction due to the SHG loss when $\beta<\chi/\sqrt{3}$, and a significantly large optical spring constant can be obtained at an intracavity power that provides $D(\bar{n})=0$.

\section{Data analysis}
Experimentally, the optical spring constants were measured via a transfer function from the applied force to the displacement of the test mass. In our experimental system, owing to photothermal effects in the nonlinear optical crystal, the displacement of the test mass did not match the cavity length, and the quadrature of the optical spring constant was rotated~\cite{PhysRevD.92.062003,Altin2017,Otabe:22}. As shown in Eqs.\,\eqref{eq:Kerrsupp-PTAbsConst} and \eqref{eq:Kerrsupp-CKopt}, the photothermal absorption rate exhibits the same functional dependence as the optical spring constant. Thus, various photothermal phenomena were enhanced by the Kerr effect. It was difficult to estimate the influence of the Kerr effect accurately using only measurements with the suspended mirror because of the large number of parameters to be estimated; therefore, we replaced the suspended mirror with a mirror equipped with a piezoelectric element (PZT) and measured the photothermal backaction~\cite{Ma:21} to compensate for the photothermal effect that complicates the optical spring measurements. To avoid changes in the Kerr gain and photothermal parameters resulting from the mirror replacement, cavity alignment was restored using only a three-axis tilt aligner equipped with a PZT mirror.\par
\begin{figure}
	\centering
	\includegraphics[width=\hsize]{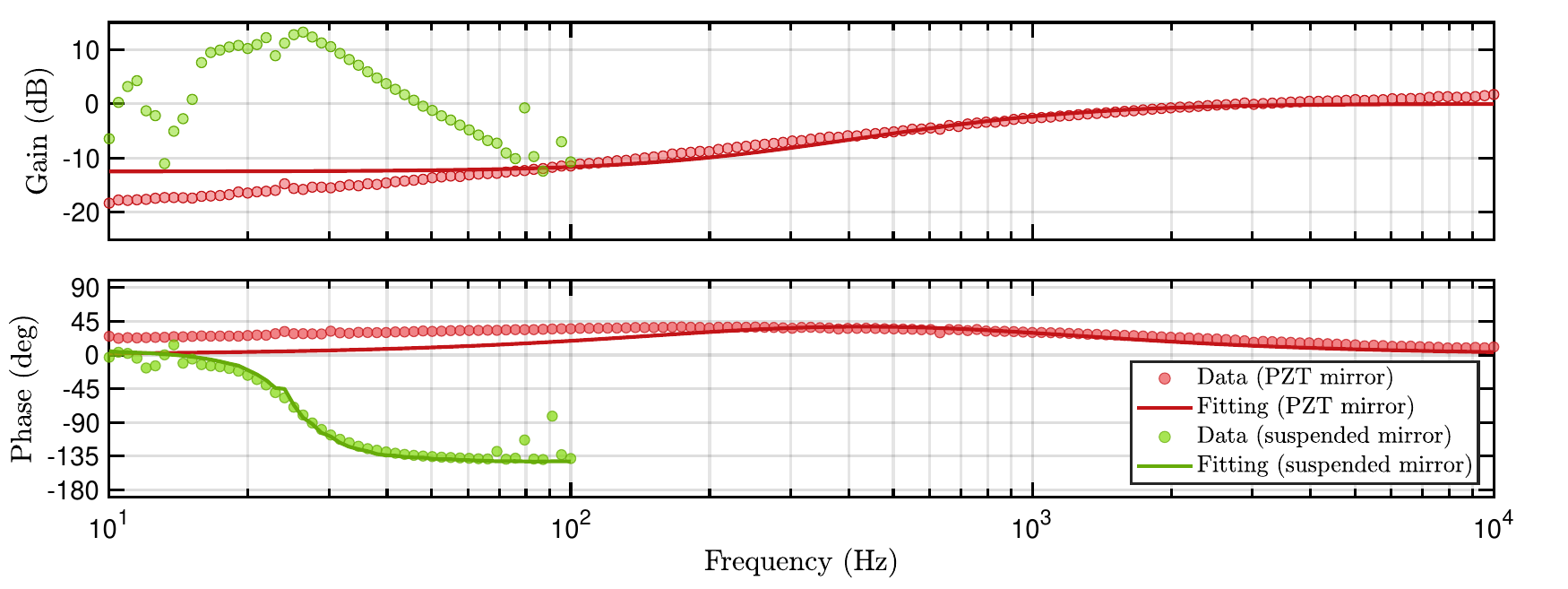}
	\caption{Example of measured transfer functions. The crystal temperature, input light power, and normalized cavity detuning were set to $T_\mathrm{C}=39.6$\,$^\circ$C, $P_0=600$\,mW, and $\xi=1.09\pm0.02$, respectively. The transfer function measurement scheme was similar to that described in Ref.~\cite{Otabe:22}. Note that the optical spring and photothermal absorption were hardly affected by the Kerr effect when the crystal temperature was set to approximately $120$\,$^\circ$C, as in the previous study. For the PZT mirror measurement, the transfer function measured with $T_\mathrm{C}=120$\,$^\circ$C, $P_0=5$\,mW, and $\xi=1.73\pm0.05$ is subtracted from the result to compensate for the frequency response of the PZT.}
	\label{fig:Kerrsupp-TF}
\end{figure}
Figure\,\ref{fig:Kerrsupp-TF} shows an example of measured transfer functions using suspended and PZT mirrors. The transfer function measured using the PZT mirror is proportional to the optical response of the cavity $H_\mathrm{th}=\delta x(\Omega)/\delta x_\mathrm{act}(\Omega)$ because the mechanical resonant frequency of the PZT is sufficiently higher than that of the optical spring. The analytical solution in the adiabatic regime ($\Omega\gg\gamma_\mathrm{th}$) is obtained from Eq.\,\eqref{eq:Kerrsupp-dxthdx} as
\begin{equation}\label{eq:Kerrsupp-Hth}
	H_\mathrm{th}(\Omega)=\dfrac{\gamma_\mathrm{th}+i\Omega}{\omega_\mathrm{th}(\Omega)+\gamma_\mathrm{th}+i\Omega}.
\end{equation}
The gain of the measured transfer function must be normalized to $1$ in the high-frequency band ($\Omega\gg\omega_\mathrm{th}$) because it depends on the conversion efficiency of the PZT and photodetector. The maximum gain of the transfer function can be estimated from the fitting; however, Eq.\,\eqref{eq:Kerrsupp-Hth} is not applicable when the adiabatic approximation is not satisfied ($\Omega<\gamma_\mathrm{th}$) because the region contributing to heat outflow is expanded. Therefore, we performed fitting using only data in the $100$--$2$\,kHz range, with an upper frequency limit defined to avoid the influence of the mechanical resonance of the PZT. The estimated photothermal absorption and relaxation rates were not used to estimate the optical spring constant because the adiabatic approximation does not hold in the $10$--$100$\,Hz band, where the transfer function was measured using a suspended mirror. In addition, the photothermal backaction must be sufficiently large ($\gamma_\mathrm{th}\lesssim\omega_\mathrm{th}$) to estimate $H_\mathrm{th}$ accurately; otherwise, the transfer function would not show significant changes due to photothermal effects. When the estimated photothermal absorption rate $\omega_\mathrm{th}$ was smaller than the photothermal relaxation rate $\gamma_\mathrm{th}$, the gain of the measured transfer function was normalized using the average value in the $6$--$7$\,kHz band, where the influence of the mechanical resonance of the PZT was negligible.\par
The transfer function measured with the suspended mirror was proportional to the susceptibility of the oscillator constrained by a combined mechanical and optical spring. When a small external force $\delta F_\mathrm{ext}(\Omega)$ is applied to the oscillator, the error signal obtained from the cavity is proportional to a small change in the cavity length $\delta x(\Omega)$. The effective susceptibility of the oscillator measured in a band sufficiently smaller than the cavity decay rate $\gamma\ (\sim10^{7}\,\mathrm{rad}/\mathrm{s})$ can be written as
\begin{equation}
	\dfrac{\delta x(\Omega)}{\delta F_\mathrm{ext}(\Omega)}=\dfrac{H_\mathrm{th}(\Omega)}{m(-\Omega^2+\Omega_\mathrm{m}^2+i\Omega\Gamma_\mathrm{m})+H_\mathrm{th}(\Omega)k_\mathrm{opt}}.
\end{equation}
The mechanical resonant frequency and Q factor were estimated to be $\Omega_\mathrm{m}/2\pi=14.0\pm0.1$\,Hz and $Q_\mathrm{m}=m\Omega_\mathrm{m}/\Gamma_\mathrm{m}=193\pm3$ by applying a sinusoidal signal and using the ring-down method. The transfer function measured using the suspended mirror shown in Fig.\,\ref{fig:Kerrsupp-TF} indicates that the oscillator is subjected to significant damping, reflecting the conversion of a positive optical spring into a positive optical damping via the photothermal effect. The influence of the photothermal effect can be described by $H_\mathrm{th}(\Omega)$, which was estimated from the measurements using the PZT mirror. By eliminating the photothermal effect, the optical spring constant $k_\mathrm{opt}$ was estimated from phase measurements using a suspended mirror. Although the optical spring constant was measured by finely varying the cavity detuning, the transfer function could not be measured within a small detuning range ($\xi\lesssim0.3$) because of the poor linearity of the Pound-Drever-Hall (PDH) signal~\cite{Drever1983}.\par
The optical spring constant is reduced by SHG and other intracavity losses, which can be estimated by measuring the ratio of the reflected light power at and outside resonance. The reflected light power $P_\mathrm{ref}$ is calculated as
\begin{equation}
	P_\mathrm{ref}=\left[1-\dfrac{4\gamma_\mathrm{in}(\gamma-\gamma_\mathrm{in})}{\gamma^2+\Delta^2}\right]P_0.
\end{equation}
By setting $T_\mathrm{C}=120$\,$^\circ$C and $P_0=5$\,mW to avoid the SHG loss and measuring the reflected light power from the cavity scan, the intracavity loss, excluding the SHG loss, was estimated to be $\gamma_\mathrm{out}/\gamma_\mathrm{in}=0.17\pm0.01$. In the setup where the optical spring constants were measured, the SHG loss was largest at resonance when we set $T_\mathrm{C}=39.6$\,$^\circ$C and $P_0=600$\,mW. The maximum value of the SHG loss was estimated to be $\gamma_\mathrm{S-max}/\gamma_\mathrm{in}=0.01\pm0.01$. Thus, the influence of the SHG loss on the intracavity power and optical spring constant was almost negligible. The lossless model is helpful for evaluating the relative magnitude of the optical spring constant because the intracavity loss, independent of the intracavity power, only changes the intracavity power and optical spring constant by a fixed factor. However, when the cavity detuning was controlled to measure the optical spring constant, thermal absorption in the nonlinear optical crystal increased the crystal temperature and changed the phase mismatch. This effect can be partially compensated for by adding an offset to the temperature control. For example, in cavity scan measurements, the transmitted light power reached a local maximum at a temperature of $T_\mathrm{C}=39.6$\,$^\circ$C, whereas this temperature decreased to $T_\mathrm{C}=38.8$\,$^\circ$C when the cavity was controlled at the resonance state using the PDH technique. The temperature at which the SHG loss reached its minimum value was assumed to be proportional to the intracavity power. The set temperature for each cavity detuning was obtained by a linear complementation between the optimum temperatures at and outside resonance.\par
Compensation achieved by changing the set temperature was insufficient. The decrease in intracavity power compared to spectrum measurements was the highest when $P_0=600$\,mW and $\xi=0$, and was $9.1\pm1.7$\% for $T_\mathrm{C}=39.6$\,$^\circ$C and $5.5\pm1.7$\% for $T_\mathrm{C}=45.4$\,$^\circ$C. We conclude that a non-negligible temperature gradient was generated in the nonlinear optical crystal, resulting in a slight shift from the phase-mismatched condition. This phenomenon can be modeled as the proportionality coefficients of the Kerr effect $\beta$ and the SHG loss $\chi$ depending on the intracavity power $P$. However, even the simplest assumption that $\beta$ and $\chi$ are represented by linear functions of $P$ leads to solving a quintic equation to determine the coefficients. Instead, we adopted a simplified model in which the maximum value of the intracavity power decreased proportionately to the intracavity power. That is, we assume that the maximum value of the transmitted light power $P_\mathrm{max}$ can be written as
\begin{equation}
	P_\mathrm{max}=\left(1-\dfrac{P_\mathrm{scan-max}}{P_\mathrm{PDH-max}}\right)P_\mathrm{trans}+P_\mathrm{scan-max},
\end{equation}
where $P_\mathrm{scan-max}$ is the maximum value of the transmitted light power measured by the cavity scan, $P_\mathrm{PDH-max}$ is the maximum value of the transmitted light power under control by the PDH technique, and $P_\mathrm{trans}$ is the transmitted light power during transfer function measurement. In this case, the normalized cavity detuning $\xi=\Delta/\gamma$, a positive value to generate a positive optical spring, can be estimated as follows:
\begin{equation}
	\xi=\sqrt{\dfrac{P_\mathrm{max}}{P_\mathrm{trans}}-1}=\sqrt{P_\mathrm{scan-max}\left(\dfrac{1}{P_\mathrm{trans}}-\dfrac{1}{P_\mathrm{PDH-max}}\right)}.
\end{equation}
This model provides a reasonable estimation when the deviation from the phase-mismatched condition and reduction in the intracavity power are small. Note that the slight deviation from the fitting observed in the optical spring estimation with $T_\mathrm{C}=39.6$\,$^\circ$C and $P_0=600$\,mW may be a systematic error resulting from incomplete modeling.\par
The optical spring constants were fitted using a lossless model formulated as follows:
\begin{equation}
	k_\mathrm{opt}=\dfrac{16}{3\sqrt{3}}k_\mathrm{opt-0}\dfrac{1}{1+\xi^2}\dfrac{\xi}{1+\xi^2+2\zeta\xi/(1+\xi^2)},
\end{equation}
where $k_\mathrm{opt-0}$ is the maximum value of the optical spring constant in the linear theory obtained with $\zeta=0$ and $\xi=1/\sqrt{3}$. $\zeta$ is the dimensionless Kerr gain, independent of the cavity detuning, and is defined as
\begin{equation}
	\zeta=(1+\xi^2)\xi_\mathrm{K}=-\dfrac{2\chi P_0}{\gamma^2\hbar\omega_0},
\end{equation}
where $\xi_\mathrm{K}=\Delta_\mathrm{K}/\gamma$ is the normalized cavity detuning owing to the Kerr effect. Even in the presence of the Kerr effect, the maximum optical spring constant is obtained at $\xi=1/\sqrt{3}$. In addition, the difference in the optical spring constant from the linear system is also maximized when $\xi=1/\sqrt{3}$. Therefore, the enhancement factor of the optical spring induced by the Kerr effect can be directly evaluated using the signal amplification ratio $A$, which is defined as 
\begin{equation}
	A=\dfrac{\left.k_\mathrm{opt}\right|_{\xi=1/\sqrt{3}}}{k_\mathrm{opt-0}}=\dfrac{1}{1-\zeta/\zeta_0},
\end{equation}
where $\zeta_0=-8/(3\sqrt{3})$ is the critical Kerr gain at which the cavity reaches the multistable state. The input power at which the optical spring constant diverges was estimated by fitting the signal amplification ratio to the incident light power. A significantly sizable optical spring constant could be realized if the condition $\zeta=\zeta_0$ is satisfied. The following factors probably determine the upper limit of the optical spring constant observed in the experiment: the SHG loss caused by strong intracavity power reduces the signal amplification ratio, high crystal temperature stability is required to maintain a fixed Kerr gain, and the intracavity power changes dramatically with the cavity length, making it difficult to control.\par
Note that when the measurement band is sufficiently lower than the photothermal absorption rate, photothermal effects inevitably occur and rotate the quadrature of the optical spring. The optical damping induced by the photothermal effect stabilizes the optomechanical system; therefore, it must be designed to leverage the advantage of the photothermal effect based on accurate modeling~\cite{Altin2017,Otabe:22}. In cases where photothermal effects are undesirable, photothermal displacement compensation by crystals exhibiting negative thermo-optic effects can be a solution.

\section{Observation of a multistable state}
\begin{figure}
	\centering
	\includegraphics[width=\hsize]{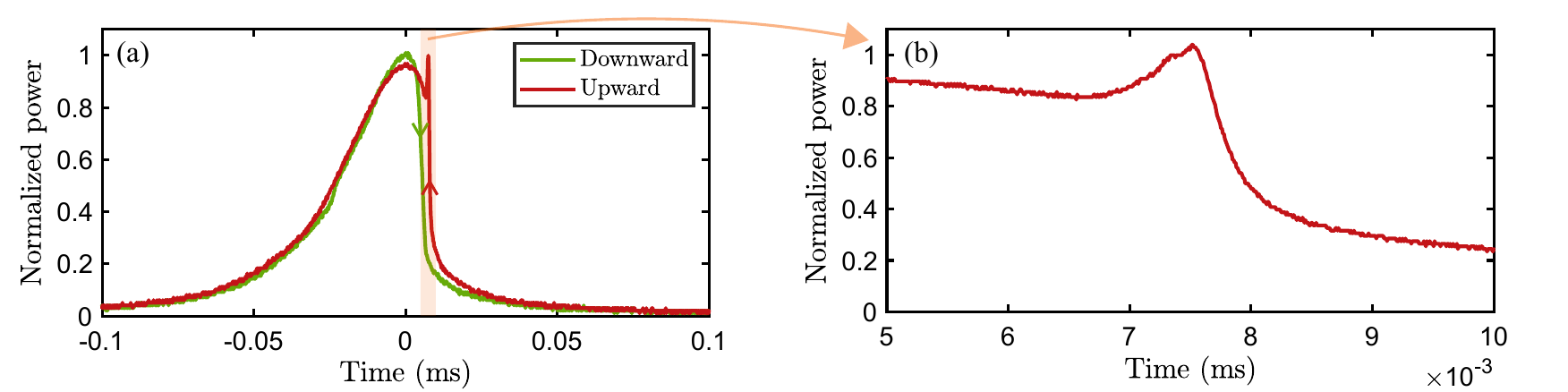}
	\caption{(a) Hysteresis measured by increasing the finesse. Downward and upward correspond to measurements with the PZT swinging in the direction of cavity length reduction and extension, respectively. Arrows added to the plot indicate the direction of the cavity scan. (b) Enlarged view of the orange-filled area of upward swing in (a).}
	\label{fig:Kerrsupp-multi}
\end{figure}
To enhance the signal amplification ratio induced by the Kerr effect, the cavity finesse must be increased. There is an exciting possibility that approaching a multistable state may induce significant signal amplification effects, potentially achieving optical spring constants of a magnitude never before observed in linear systems. In addition, realizing the critical Kerr effect, which corresponds to the transition to multistability, is essential for achieving high squeezing levels~\cite{PhysRevA.84.033839}. Experimental evidence of a multistable state has been observed as a hysteresis in the cavity spectrum~\cite{AGWhite_1996,PhysRevA.80.053801}.\par
In this study, the optical spring constants were measured using an input mirror with $94\pm1$\% power transmissivity. By replacing the input mirror with one with $98\pm1$\% power transmissivity, the finesse of the cavity was increased to $300\pm30$. Hysteresis was confirmed by measuring the spectra from a fast cavity scan using a PZT, in which the photothermal effect was negligible [Fig.\,\figref{fig:Kerrsupp-multi}{(a)}]. The Kerr effect reduced the effective cavity length as the intracavity power increased. The downward swing exhibited a relatively gradual change in intracavity power because the effective cavity length underwent negative feedback after exceeding the resonance state. In contrast, the upward swing exhibited a rapid change in the transmitted light power. The rise time was independent of the swing speed and equivalent to the cavity charging time $\tau=2\pi/\gamma$ ($\sim10^{-6}$\,s), as shown in Fig.\,\figref{fig:Kerrsupp-multi}{(b)}. The peak at approximately $7.5$\,\textmu s may be an overshoot of the photodetector. An optical spring with a remarkably high resonant frequency can be generated by controlling the appropriate cavity detuning. However, because of the instability caused by the drastic rate of change in the cavity length, the cavity length could not be controlled. Stiffer optical springs can be obtained by improving the control mechanism and achieving stable control within a narrow cavity length range.

\section{Correspondence between OPA and Kerr schemes}
The enhancement of an optical spring achieved by inducing optical parametric amplification (OPA) in the signal recycling cavity of a gravitational wave detector has been theoretically investigated~\cite{SOMIYA2016521,KOROBKO20182238,PhysRevD.107.122005}. The principle of this method is active amplification of the gravitational wave signal by nonlinear optical effects, which can also be achieved by inducing the Kerr effect in the arms~\cite{PhysRevLett.95.193001}. However, it is not apparent that the Kerr effect provides the same signal amplification effect as OPA when the signal recycling cavity is detuned to generate an optical spring. This section presents the correspondence between the OPA and Kerr schemes and discusses the characteristics of the Kerr system.\par
\begin{figure}
	\centering
	\includegraphics[width=0.5\hsize]{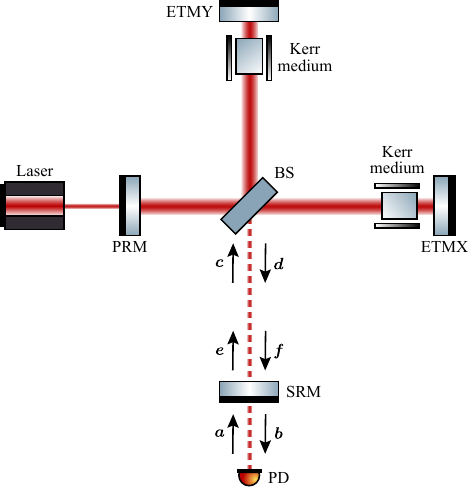}
	\caption{Schematic of the gravitational wave signal amplification system based on the Kerr scheme. The laser light amplified by the power recycling mirror (PRM) is split by an isotropic beam splitter (BS) and propagates through the arms containing the Kerr medium. Gravitational waves produce differential displacements in the end test masses (ETMs) X and Y, and the differential signal is amplified by the signal recycling mirror (SRM). A photodetector (PD) at the dark port detects the gravitational wave signal.}
	\label{fig:Kerrsupp-KerrGWD}
\end{figure}
The two-photon formalism~\cite{PhysRevA.31.3068} is helpful in calculating the sensitivity of gravitational wave detectors. Let $a_1(\Omega)$ and $a_2(\Omega)$ be the amplitude and phase quadratures of the input light-field fluctuations from the dark port, respectively. Herein, we discuss a lossless dual-recycling Michelson interferometer with Kerr medium in the arms [Fig.\,\ref{fig:Kerrsupp-KerrGWD}]. By calculating the transformation that an interferometer controlled at the dark fringe gives to the incident light field $\bm{a}=(a_1,\ a_2)^\mathrm{T}$, the input-output relation can be written as
\begin{equation}
	\bm{b}=-r_\mathrm{s}\bm{a}+t_\mathrm{s}\bm{f},\ \ \ \bm{e}=r_\mathrm{s}\bm{f}+t_\mathrm{s}\bm{a},\ \ \ \bm{c}=\mathbb{R}(\phi)\bm{e},\ \ \ \bm{f}=\mathbb{R}(\phi)\bm{d},\ \ \ \bm{d}=\mathbb{R}(\Phi)\mathbb{K}(-2\Phi)\left[\mathbb{K}(\mathcal{K})\bm{c}\mathrm{e}^{2i\beta}+\alpha\bm{h}\mathrm{e}^{i\beta}\right],
\end{equation}
where $r_\mathrm{s}^2$ and $t_\mathrm{s}^2$ are the power reflectivity and transmissivity of the signal recycling mirror, respectively, $\beta=L_\mathrm{arm}\Omega/c$ with arm length $L_\mathrm{arm}$ is the phase delay per single trip in the arm, $\phi$ is the detune phase of the signal recycling cavity, $\Phi=-4L_\mathrm{arm}^2\chi P_\mathrm{arm}/(\hbar\omega_0 c^2)$ with intra-arm power $P_\mathrm{arm}$ is the phase change induced by the Kerr effect, $\mathcal{K}=8\omega_0 P_\mathrm{arm}/(mc^2\Omega^2)$ is the coupling constant of the light field and radiation pressure force, $\alpha=\sqrt{4\omega_0 P_\mathrm{arm}L_\mathrm{arm}^2/(\hbar c^2)}$ is the signal strength, and $\bm{h}=(0,\ h)^\mathrm{T}$ with $h$ the gravitational wave signal in the strain. $\mathbb{R}(\phi)$ and $\mathbb{K}(\mathcal{K})$ denote the rotation and ponderomotive squeezing matrices, respectively:
\begin{equation}
	\mathbb{R}(\phi)=\begin{pmatrix}
		\cos\phi&-\sin\phi\\\sin\phi&\cos\phi
	\end{pmatrix},\ \ \ \mathbb{K}(\mathcal{K})=\begin{pmatrix}
	1&0\\-\mathcal{K}&1
\end{pmatrix}.
\end{equation}
The transformation of the Kerr effect on the light field can be represented by the product of the rotation and ponderomotive squeezing matrices~\cite{PhysRevLett.95.193001}, which can be rewritten as
\begin{equation}
	\mathbb{R}(\Phi)\mathbb{K}(-2\Phi)=\mathbb{R}(\Phi)\mathbb{S}(\exp(\mathrm{arcsinh}(\Phi)),-\mathrm{arccot}(\Phi)/2)\mathbb{R}(\arctan(\Phi))\ \ \ (\mathrm{for}\ \Phi<0),
\end{equation}
where $\mathbb{S}(s,\eta)$ with a squeezing factor $s$ and squeezing angle $\eta$ denotes the squeezing matrix:
\begin{equation}
	\mathbb{S}(s,\eta)=\mathbb{R}(\eta)\begin{pmatrix}
		s&0\\0&1/s
	\end{pmatrix}\mathbb{R}(-\eta).
\end{equation}
\par
In a recent study~\cite{PhysRevD.107.122005}, we investigated a signal amplification system based on the OPA scheme in which the squeezing factor and squeezing angle of OPA were $s$ and $\eta$, respectively. A comparison of the calculations shows that the Kerr scheme is equivalent to the OPA scheme with the parameters transformed as follows:
\begin{align}
	s&\rightarrow\exp(\mathrm{arcsinh}(\Phi)),\\
	\eta&\rightarrow-\mathrm{arccot}(\Phi)/2,\\
	\phi&\rightarrow\phi+\Phi.
\end{align}
Here, we assume that the phase change induced by the Kerr effect is sufficiently small, such that $\arctan(\Phi)\simeq\Phi$. The Kerr scheme provides the same effect on the interferometer as the OPA scheme with a fixed squeezing angle. The Kerr system does not allow techniques requiring a variable squeezing angle, such as dynamic tuning of the interferometer response~\cite{KOROBKO20182238}; therefore, it is also essential to investigate the experimental system based on the OPA scheme. Conversely, implementing the Kerr scheme does not add degrees of freedom that must be controlled and does not require an additional control light field or complex control mechanism as are required for the OPA scheme, which is a clear advantage of the Kerr scheme. The Kerr scheme is one of the most promising techniques for improving the sensitivity of gravitational wave detectors in the high-frequency band.\par
The reflected light field from the interferometer can be calculated as
\begin{equation}
	\bm{b}=\dfrac{1}{M}\left(\mathbb{A}\bm{a}\mathrm{e}^{2i\beta}+\mathbb{H}\bm{h}\mathrm{e}^{i\beta}\right),
\end{equation}
with
\begin{align}
	M&=1+r_\mathrm{s}^2\mathrm{e}^{4i\beta}-r_\mathrm{s}\mathrm{e}^{2i\beta}\left[2\cos(2\phi+\Phi)+(\mathcal{K}-2\Phi)\sin(2\phi+\Phi)\right],\\
	A_{11}&=(1+r_\mathrm{s}^2)\cos(2\phi+\Phi)+\dfrac{1}{2}(\mathcal{K}-2\Phi)[t_\mathrm{s}^2\sin\Phi+(1+r_\mathrm{s}^2)\sin(2\phi+\Phi)]-2r_\mathrm{s}\cos(2\beta),\\
	A_{12}&=-t_\mathrm{s}^2[(\mathcal{K}-2\Phi)\sin\phi\sin(\phi+\Phi)+\sin(2\phi+\Phi)],\\
	A_{21}&=-t_\mathrm{s}^2[(\mathcal{K}-2\Phi)\cos\phi\cos(\phi+\Phi)-\sin(2\phi+\Phi)],\\
	A_{22}&=(1+r_\mathrm{s}^2)\cos(2\phi+\Phi)+\dfrac{1}{2}(\mathcal{K}-2\Phi)[-t_\mathrm{s}^2\sin\Phi+(1+r_\mathrm{s}^2)\sin(2\phi+\Phi)]-2r_\mathrm{s}\cos(2\beta),\\
	H_{12}&=-t\alpha[r_\mathrm{s}\mathrm{e}^{2i\beta}\sin\phi+\sin(\phi+\Phi)],\\
	H_{22}&=t\alpha[-r_\mathrm{s}\mathrm{e}^{2i\beta}\cos\phi+\cos(\phi+\Phi)],\\
	H_{11}&=H_{21}=0.
\end{align}
The resonant angular frequency of the optical spring $\Omega_\mathrm{opt}$ is obtained by solving the equation $M=0$ with $\beta=0$~\cite{PhysRevD.65.042001}. The optical spring constant $k_\mathrm{opt}=m\Omega_\mathrm{opt}^2$ can be written as
\begin{align}\label{eq:Kerrsupp-koptSRMI}
 	k_\mathrm{opt}&=\dfrac{8\omega_0 P_\mathrm{arm}}{c^2}\dfrac{\sin(2\phi+\Phi)}{r_\mathrm{s}+1/r_\mathrm{s}-2\cos(2\phi+\Phi)+2\Phi\sin(2\phi+\Phi)}\nonumber\\
 	&\simeq\dfrac{4\omega_0 P_\mathrm{arm}}{L_\mathrm{arm}c}\dfrac{\Delta}{\gamma^2+\Delta^2+2\Delta\Delta_\mathrm{K}}.
\end{align}
Here, we ignore terms above the third order in the microquantities $t_\mathrm{s}^2$, $\Phi$, and $\phi$, and redefine $\gamma=t_\mathrm{s}^2c/(4L_\mathrm{arm})$, $\Delta_\mathrm{K}=\Phi c/(2L_\mathrm{arm})$, and $\Delta=\phi c/L_\mathrm{arm}+\Delta_\mathrm{K}$. By normalizing with intracavity power, Eq.\,\eqref{eq:Kerrsupp-koptSRMI} is consistent with Eq.\,\eqref{eq:Kerrsupp-kopt} in the lossless case because the input-output relation in the Fabry-Perot cavity is equivalent to that in the dual-recycling Michelson interferometer. We observed that the Kerr effect enhanced the optical spring in the Fabry-Perot cavity; this result indicates that the same enhancement of optomechanical coupling can be realized in a gravitational wave detector based on the Kerr scheme.

\end{document}